\documentclass{aastex62}

\received{...}
\revised{...}
\accepted{...}

\submitjournal{ApJL}

\shorttitle{Coronal condensation by magnetic reconnection}
\shortauthors{Li et al.}

\begin{document}

\title{Coronal condensations caused by magnetic reconnection between solar coronal loops}

\correspondingauthor{Leping Li}
\email{lepingli@nao.cas.cn}

\author[0000-0002-0786-7307]{Leping Li}
\affil{CAS Key Laboratory of Solar Activity, National Astronomical Observatories, Chinese Academy of sciences, Beijing 100101, China}
\affiliation{Shandong Provincial Key Laboratory of Optical Astronomy and Solar-Terrestrial Environment, and Institute of Space Sciences, Shandong University, Weihai, Shandong 264209, China}
\affiliation{University of Chinese Academy of Sciences, Beijing 100049, China}

\author{Jun Zhang}
\affil{CAS Key Laboratory of Solar Activity, National Astronomical Observatories, Chinese Academy of sciences, Beijing 100101, China}
\affiliation{University of Chinese Academy of Sciences, Beijing 100049, China}

\author{Hardi Peter}
\affiliation{Max Planck Institute for Solar System Research, 37077 G\"{o}ttingen, Germany}

\author{Lakshmi Pradeep Chitta}
\affiliation{Max Planck Institute for Solar System Research, 37077 G\"{o}ttingen, Germany}

\author{Jiangtao Su}
\affil{CAS Key Laboratory of Solar Activity, National Astronomical Observatories, Chinese Academy of sciences, Beijing 100101, China}
\affiliation{University of Chinese Academy of Sciences, Beijing 100049, China}

\author{Chun Xia}
\affiliation{School of Physics and Astronomy, Yunnan University, Kunming 650050, China}
\affiliation{Centre for mathematical Plasma Astrophysics, Department of Mathematics, KU Leuven, Celestijnenelaan 200B, B-3001 Leuven, Belgium}
\affil{CAS Key Laboratory of Solar Activity, National Astronomical Observatories, Chinese Academy of sciences, Beijing 100101, China}

\author{Hongqiang Song}
\affiliation{Shandong Provincial Key Laboratory of Optical Astronomy and Solar-Terrestrial Environment, and Institute of Space Sciences, Shandong University, Weihai, Shandong 264209, China}
\affil{CAS Key Laboratory of Solar Activity, National Astronomical Observatories, Chinese Academy of sciences, Beijing 100101, China}

\author{Yijun Hou}
\affil{CAS Key Laboratory of Solar Activity, National Astronomical Observatories, Chinese Academy of sciences, Beijing 100101, China}
\affiliation{University of Chinese Academy of Sciences, Beijing 100049, China}

\begin{abstract}

Employing Solar Dynamics Observatory (SDO)/Atmospheric Imaging Assembly (AIA) multi-wavelength images, we report the  coronal condensation during the magnetic reconnection (MR) between a system of open and closed coronal loops. Higher-lying magnetically open structures, observed in AIA 171\,\AA~images above the solar limb, move downward and interact with the lower-lying closed loops, resulting in the formation of dips in the former. An X-type structure forms at the interface. The interacting loops reconnect and disappear. Two sets of newly-reconnected loops then form and recede from the MR region. During the MR process, bright emission appears sequentially in the AIA 131\,\AA~and 304\,\AA~channels repeatedly in the dips of higher-lying open structures. This indicates the cooling and condensation process of hotter plasma from $\sim$0.9 MK down to $\sim$0.6 MK, and then to $\sim$0.05 MK, also supported by the light curves of the AIA 171\,\AA, 131\,\AA, and 304\,\AA~channels. The part of higher-lying open structures supporting the condensations participate in the successive MR. The condensations without support by underlying loops then rain back to the solar surface along the newly-reconnected loops. Our results suggest that the MR between coronal loops leads to the condensation of hotter coronal plasma and its downflows. MR thus plays an active role in the mass cycle of coronal plasma because it can initiate the catastrophic cooling and condensation. This underlines that the magnetic and thermal evolution has to be treated together and cannot be separated, even in the case of catastrophic cooling.

\end{abstract}


\section{Introduction} \label{sec:int}

Magnetic reconnection (MR), the reconfiguration of magnetic field geometry, is a fundamental process in magnetized plasma systems throughout the universe, such as accretion disks, solar and stellar coronae, planetary magnetospheres, and laboratory plasmas \citep{prie00}. It is considered to play an essential role in the rapid release of magnetic energy and its conversion to other forms, e.g. thermal, kinetic, and particles \citep{prie86,prie00}. In solar physics, numerous theoretical concepts of MR have been used to explain various features, e.g. flares, filament eruptions, and coronal mass ejections \citep{shib99,lin00,mei17}. In two-dimensional models, MR occurs at an X-point where anti-parallel magnetic field lines converge and reconnect \citep{prie86,prie00}. The process of MR is difficult to observe directly. However, coronal structures, e.g. loops, and their structural changes often outline the magnetic field topology and its evolution, because the magnetic flux is frozen into the coronal plasma \citep{su13,li16a}. So far, using remote-sensing observations, many MR signatures have been reported, e.g. cusp-shaped post-flare loops \citep{tsun92,yan18}, loop-top hard X-ray sources \citep{masu94,su13}, MR inflows \citep{yoko01,li09,yang15,huang18} and outflows \citep{taka12,tian14,li16c,ning16}, supra-arcades downflows \citep{mcke00, inne03, sava12, li16b}, current sheets \citep{liu10,kwon16,li16a,li16b,li18,xue18}, and plasmoid ejections \citep{kuma13,yang17,yang18,zheng17}.

The condensation of cool plasma out of the hot corona is a widely observed phenomenon best seen at the solar limb. One widely investigated concept for this process is based on the thermal properties of the plasma alone \citep{mull03}. It is independent of the (evolution of the) coronal magnetic field, and only the loss of equilibrium between heat input, heat conduction and radiative losses causes the plasma to cool catastrophically \citep{xia16}. In numerical models, thermal instabilities occurring within a current sheet and an accompanying MR result in the formation of a quiescent (current sheet) prominence \citep{smith77,malh83}. On the other hand, MR can create a helical flux rope, and the cooling radiation and condensation of plasma trapped inside the flux rope form the cool dense plasma of prominences \citep{pneu83,kane15}. Recently, \citet{kane17} propose a MR-condensation prominence formation model, and demonstrate that MR can lead to flux rope formation and radiative condensation under certain conditions. Employing Solar Dynamics Observatory \citep[SDO,][]{pesn12}/Atmospheric Imaging Assembly \citep[AIA,][]{lemen12} images, the cooling and condensation of coronal plasma are observed in a loop system \citep{liu12} and a coronal cavity \citep{berg12} for prominence formations, and in an active region loop \citep{vash15} for coronal rain formation, respectively.

In this Letter, we study the MR between coronal loops, and firstly report the cooling and condensation of coronal plasma during the MR. The observations and results are presented in Sections\,\ref{sec:obs} and \ref{sec:res}, and a summary and discussion is presented in Section\,\ref{sec:sum}.

\section{Observations}\label{sec:obs}

AIA is a set of normal incidence imaging telescopes, acquiring solar atmospheric images in ten wavelength bands. Different AIA channels show plasma at different temperatures, e.g. 171\,\AA~peaks at $\sim$0.9 MK (Fe IX), 131\,\AA~peaks at $\sim$0.6 MK (Fe VIII) and $\sim$10 MK (Fe XXI), and 304\,\AA~peaks at $\sim$0.05 MK (He II). In this study, we employ AIA 171\,\AA, 131\,\AA, and 304\,\AA~images, with spatial sampling and time cadence of 0.6\arcsec\,pixel$^{-1}$ and 12\,s, to investigate the evolution of MR between loops and coronal condensation.

\section{Results}\label{sec:res}

On January 19, 2012, above the northwestern solar limb, sets of loops were observed in AIA 171\,\AA~images. The field of view (FOV) of the region-of-interest with respect to the full Sun is denoted by the red rectangle on the white circle in Figure\,\ref{f:reconnection}(a). For better displaying the evolution of loops, AIA images are rotated counter-clockwise by an angle of 40$^{\circ}$. Thus the portion of the limb in the region-of-interest is roughly horizontal in the images presented here.

\subsection{MR between coronal loops}

\begin{figure}[ht!]
\centering
\plotone{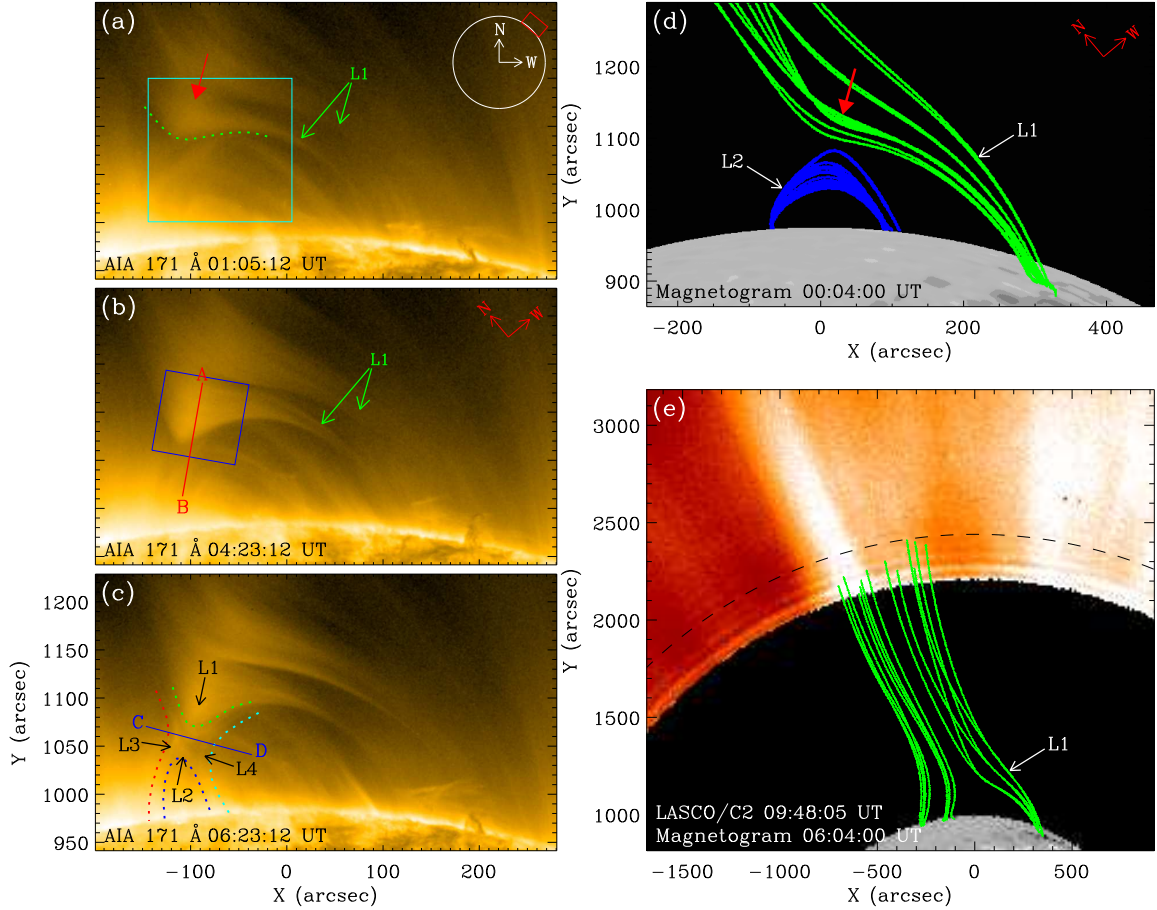}
\caption{Evolution of the coronal emission and magnetic field lines from a PFSS model. ((a)-(c)) AIA 171 \AA~images. ((d)-(e)) Coronal magnetic field lines derived from the PFSS model. The red rectangle on the white circle in (a) indicates the location of FOV of ((a)-(c)) with respect to the full Sun. The AIA images are rotated counter-clockwise by an angle of 40$^{\circ}$. The white and red arrows N and W in (a), (b), and (d) show the north and west directions in the FOVs of AIA before and after the rotation, respectively. The cyan rectangle in (a) shows the FOV of Figure\,\ref{f:condensation}, and the blue rectangle in (b) marks the region for the light curve of the AIA 171\,\AA~channel as displayed in Figure\,\ref{f:signatures}(a) by the blue line. The green, blue, red, and cyan dotted lines in (a) and (c) outline the coronal loops L1, L2, L3, and L4, and the red solid arrows in (a) and (d) mark a dip of loops L1. The red and blue lines AB and CD in (b) and (c) separately indicate the positions of time-slices of AIA 171\,\AA~images as displayed in Figures\,\ref{f:signatures}(a) and (b). In (d) and (e), the blue and green lines represent the closed and open PFSS coronal magnetic field lines at 00:04:00 UT (d) and 06:04:00 UT (e), respectively, and the inner gray-scale images show the line of sight (LOS) magnetograms in a range of $\pm$100 Mx cm$^{-2}$. In (e), the outer image displays the LASCO/C2 coronagraph image at 09:48:05 UT, and the dashed line describes the upper radial boundary of the PFSS model corona, also called the source surface, at 2.5 solar radii. 
\label{f:reconnection}}
\end{figure}

The loops L1, show curved open structures with a dip, denoted by a red arrow, at a height of $\sim$125\,Mm above the limb, see Figure\,\ref{f:reconnection}(a). Since $\sim$01:00 UT, the dip of loops L1 moves down toward the solar surface, becoming deeper, see Figure\,\ref{f:reconnection}(b), and interacts with the underlying closed loops L2, see Figure\,\ref{f:reconnection}(c). An X-type structure is formed at the interface of these two loops, enclosed by the green and blue dotted lines in Figure\,\ref{f:reconnection}(c). The interacting loops L1 and L2 disappear, and two sets of new loops L3 and L4, marked by the red and cyan dotted lines in Figure\,\ref{f:reconnection}(c), form and retract away from the interacting region horizontally. This evolution of loops moving in (L1 and L2) and out (L3 and L4) of an X-type configuration well captures the ongoing MR. The MR continues until  $\sim$23:00 UT, lasting for $\sim$22\,hr (see Animation 1).

Employing the potential field source surface (PFSS) model\footnote{More information of PFSS model we employed can be found at \url{http://www.lmsal.com/~derosa/pfsspack/}} \citep{scha69} which provides an approximate description of the solar coronal magnetic field based on the observed photospheric magnetic fields (magnetograms), we derive the coronal magnetic field in a spherical shell spanning radial distances from 1.0 to 2.5 solar radii. The green and blue lines in Figure\,\ref{f:reconnection}(d) show the selected open and closed field lines. They clearly resemble the observed loops L1 (open) and L2 (closed) in Figures\,\ref{f:reconnection}((a)-(c)), supporting that the magnetic flux is frozen into the plasma trapped in the coronal structures. According to the PFSS coronal field, the loops L1 root in the negative polarity magnetic concentrations, and the loops L2 connect the right positive and left negative polarity magnetic concentrations, in the quiet Sun. From the combination of the PFSS field lines and the LASCO/C2 coronagraph image in Figure\,\ref{f:reconnection}(e), we see bright streamer-like structures in the outer corona, consistent with the PFSS open field lines (see the green lines in Figure\,\ref{f:reconnection}(e)). This further supports that the loops L1 show open coronal field lines.

\begin{figure}[ht!]
\plotone{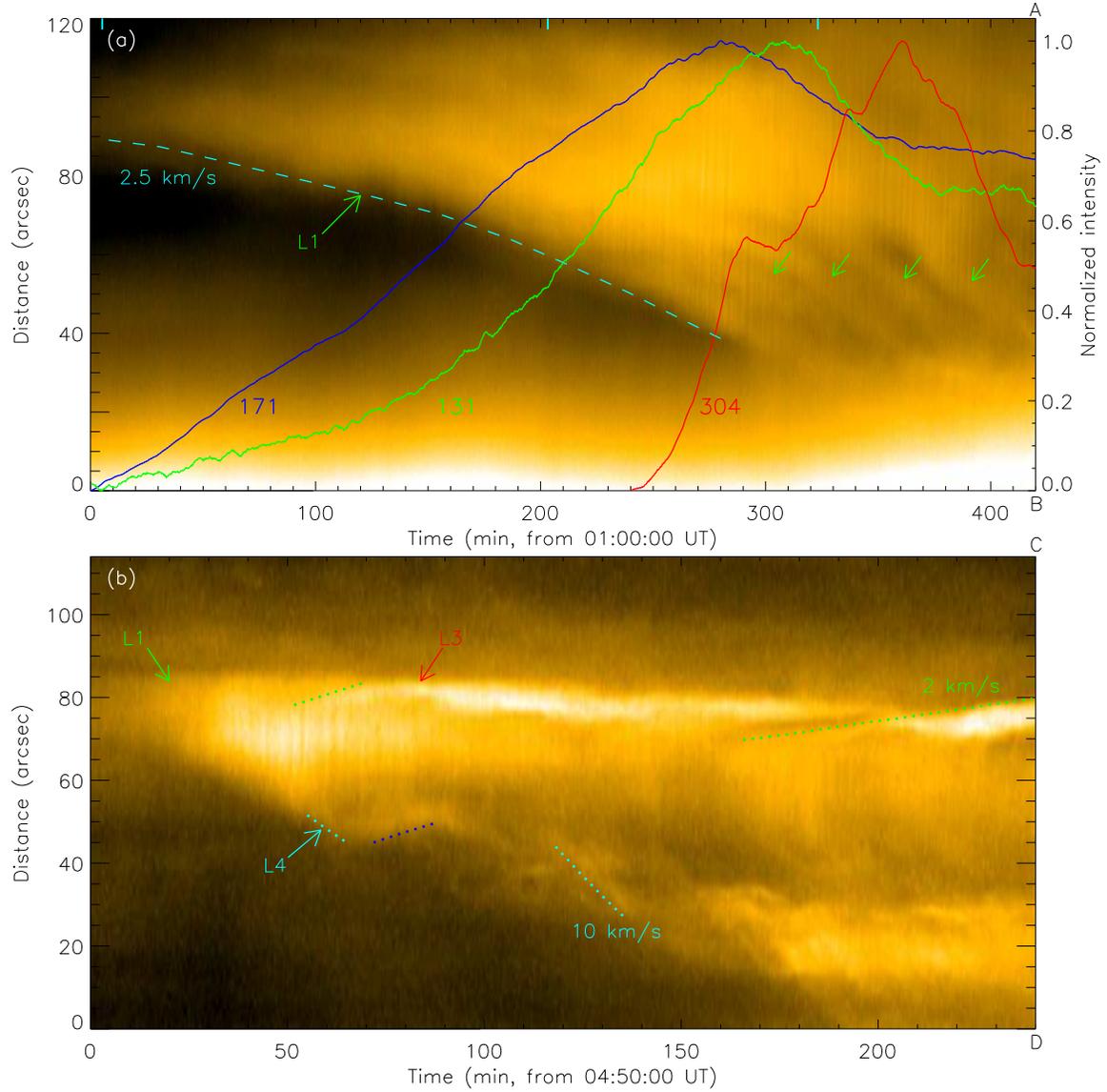}
\centering
\caption{Temporal evolution of the coronal emission near the MR region. Time-slices of AIA 171\,\AA~images along the red and blue lines AB (a) and CD (b) in Figures\,\ref{f:reconnection}(b) and (c), respectively. In (a), the blue, green, and red lines separately show the light curves of the AIA 171\,\AA, 131\,\AA, and 304\,\AA~channels in the blue and yellow rectangles in Figures\,\ref{f:reconnection}(b), \ref{f:condensation}(e), and \ref{f:condensation}(i). The cyan vertical lines denote the times of Figures\,\ref{f:reconnection}((a)-(c)). The green arrows mark the fine structures of loops L1. The cyan dashed line outlines the downward motion of loops L1. In (b), the green dotted lines show the contractions of loops L3, and the cyan dotted lines outline the contractions of loops L4. The blue dotted line denotes the backward motion of loops L4. The moving speeds of loops L1, L3, and L4 are denoted by the numbers in ((a)-(b)).
\label{f:signatures}}
\end{figure}

Along the red line AB in Figure\,\ref{f:reconnection}(b), a time-slice of AIA 171\,\AA~images is made, and displayed in Figure\,\ref{f:signatures}(a). It indicates that the loops L1 move slowly with mean speeds of 2-7 km\,s$^{-1}$, and accelerations of 0.1-0.8 m\,s$^{-2}$. Moreover, several fine structures of loops L1 are detected, marked by the green arrows in Figure\,\ref{f:signatures}(a). Along the blue line CD in Figure\,\ref{f:reconnection}(c), another time-slice of AIA 171\,\AA~images is made, and shown in Figure\,\ref{f:signatures}(b). The retractions of the newly-reconnected loops L3 and L4, marked by green and cyan dotted lines in Figure\,\ref{f:signatures}(b), are detected away from the MR region between loops L1 and L2, with mean speeds of $\sim$2 km\,s$^{-1}$ and $\sim$10 km\,s$^{-1}$, respectively. Moreover, the loops L4 move backward, perturbed by the neighboring filament eruption, with a mean speed of 4 km\,s$^{-1}$, denoted by the blue dotted line in Figure\,\ref{f:signatures}(b).

In the blue rectangle, enclosing the dip region of loops L1, in Figure\,\ref{f:reconnection}(b), the light curve of the AIA 171\,\AA~channel is calculated, and displayed in Figure\,\ref{f:signatures}(a) as a blue line. It increases first, reaches the peak at $\sim$05:40 UT, and then decreases slowly, concurrent with the downward motion of loops L1 and the MR between loops L1 and L2.

\subsection{Coronal condensation}

\begin{figure}[ht!]
\plotone{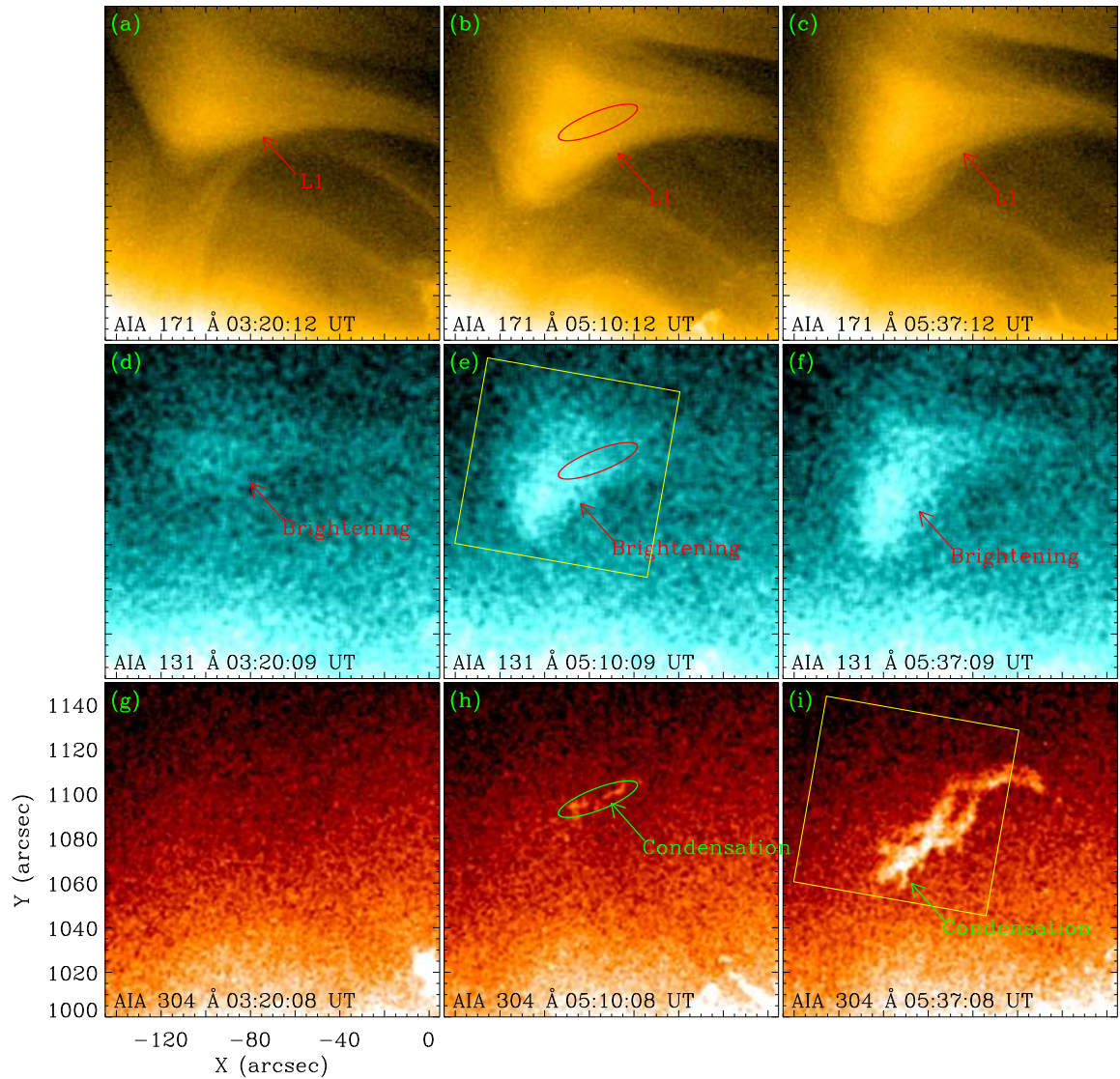}
\centering
\caption{Evolution of the condensation from $\sim$1\,MK to below 0.1\,MK. AIA 171\,\AA~((a)-(c)), 131\,\AA~((d)-(f)), and 304\,\AA~((g)-(i)) images. The yellow rectangles in (e) and (i) mark the regions for the light curves of the AIA 131\,\AA~and 304\,\AA~channels as displayed in Figure\,\ref{f:signatures}(a) by the green and red lines, respectively. The red and green ellipses in (b), (e), and (h) enclose the condensation of coronal plasma in (h). 
\label{f:condensation}}
\end{figure}

Along with the downward motion of the dip of AIA 171\,\AA~loops L1, see Figures\,\ref{f:condensation}((a)-(c)), a bright emission appears in AIA 131\,\AA~images since $\sim$02:00 UT in the dip region, see Figures\,\ref{f:condensation}((d)-(f)). As there is no associated observation of the bright emission in AIA higher-temperature channels, e.g. 193\,\AA, 211\,\AA, 335\,\AA, and 94\,\AA, the AIA 131\,\AA~emission thus shows the cooler ($\sim$0.6 MK) rather than the hotter ($\sim$10 MK) plasma, cooler than the emitting plasma detected by the AIA 171\,\AA~channel ($\sim$0.9\,MK). Together with the AIA 171\,\AA~loops L1, the AIA 131\,\AA~emission also moves downward with similar speeds (see Animation 1). In the yellow rectangle in Figure\,\ref{f:condensation}(e), the light curve of the AIA 131\,\AA~channel is calculated, and displayed in Figure\,\ref{f:signatures}(a) as a green line. It first increases, reaches the peak, and then slowly decreases, similar to that of AIA 171\,\AA~channel (the blue line). However, it peaks at $\sim$06:10 UT,  $\sim$30\,min after the peak of AIA 171\,\AA~light curve.

Since $\sim$05:05 UT, a bright condensation appears in AIA 304\,\AA~channel that shows much cooler ($\sim$0.05\,MK) plasma, see Figures\,\ref{f:condensation}((g)-(i)). It first appears at the right edge, rather than the bottom, of the dip region, see the ellipses in Figures\,\ref{f:condensation}(b), (e), and (h), and then extends to both sides along the loops L1 with a mean speed of $\sim$8 km\,s$^{-1}$. In the yellow rectangle in Figure\,\ref{f:condensation}(i), the light curve of the AIA 304\,\AA~channel is measured, and shown in Figure\,\ref{f:signatures}(a) as a red line. Similar to those in the AIA 171\,\AA~and 131\,\AA~channels, it increases quickly after its appearance, reaches the peak, and then decreases. However, it peaks at $\sim$07:00 UT, $\sim$50 (80)\,min later than the peak of AIA 131 (171)\,\AA~light curve.

The images and light curves of the AIA 171\,\AA, 131\,\AA, and 304\,\AA~channels clearly show the cooling and condensation process of hotter coronal plasma in the dip region of loops L1. This indicates that the coronal plasma cools down from $\sim$0.9 MK, the characteristic temperature of AIA 171\,\AA~channel, to $\sim$0.6 MK, the characteristic temperature of AIA 131\,\AA~channel, in $\sim$30 min, and then to $\sim$0.05 MK, the characteristic temperature of AIA 304\,\AA~channel, in $\sim$50 min.

\subsection{Condensation downflows}

\begin{figure}[ht!]
\centering
\includegraphics[width=0.5\textwidth]{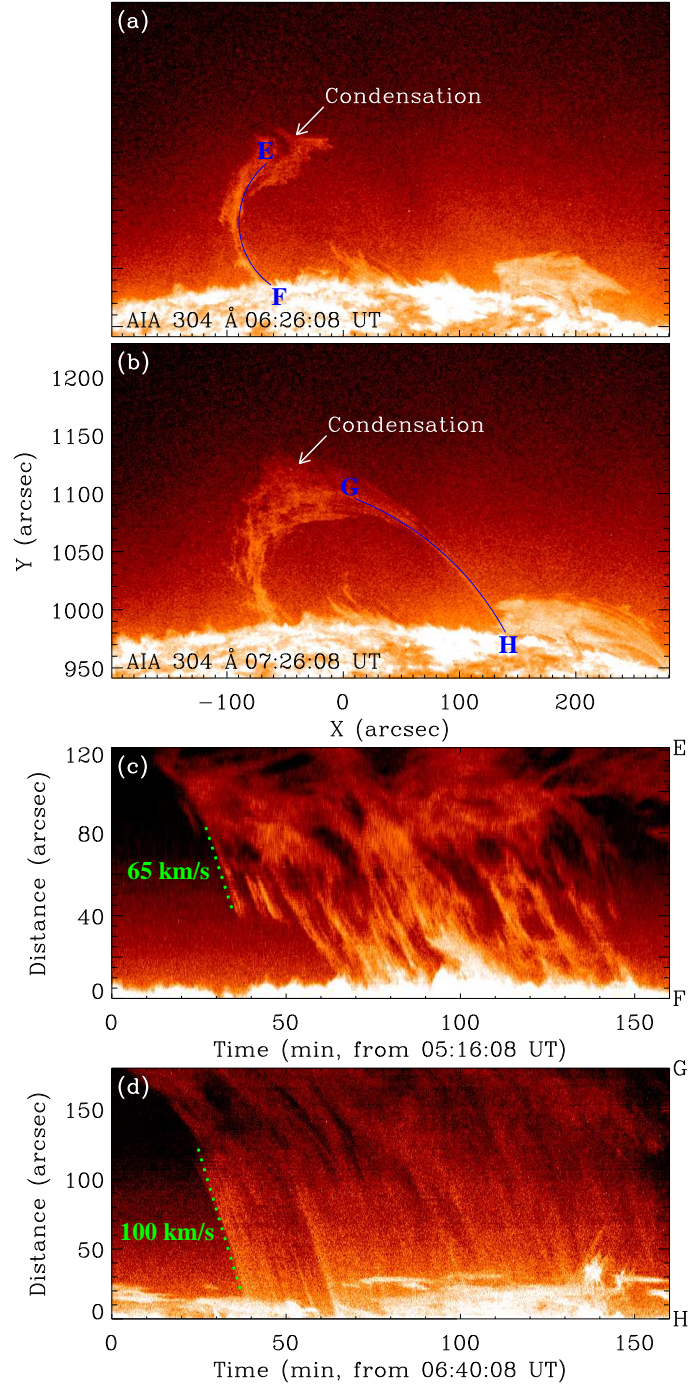}
\caption{Temporal evolution of the condensation. ((a)-(b)) AIA 304\,\AA~images. ((c)-(d)) Time-slices of AIA 304\,\AA~images separately along the blue lines EF (c) and GH (d) in (a) and (b). The green dotted lines in (c) and (d) outline the downflows of the condensations. 
\label{f:downflow}}
\end{figure}

The condensation appears and moves higher away from the lower edge of loops L1, see the red ellipse in Figure\,\ref{f:condensation}(b). Along with the successive MR between loops L1 and L2, the part of field lines of loops L1, supporting the condensation, take part in the MR. Newly-reconnected loops L3 and L4 then form. Lacking the support by lower-lying loops L1, since $\sim$05:47 UT, the condensation flows toward the chromosphere first along the left leg of loops L4, see Figure\,\ref{f:downflow}(a). Along the blue line EF in Figure\,\ref{f:downflow}(a), we make a time-slice of AIA 304\,\AA~images, and show it in Figure\,\ref{f:downflow}(c). Multiple downflows are detected, with a mean speed of 60 km\,s$^{-1}$.

\begin{figure}[ht!]
\centering
\includegraphics[width=\textwidth]{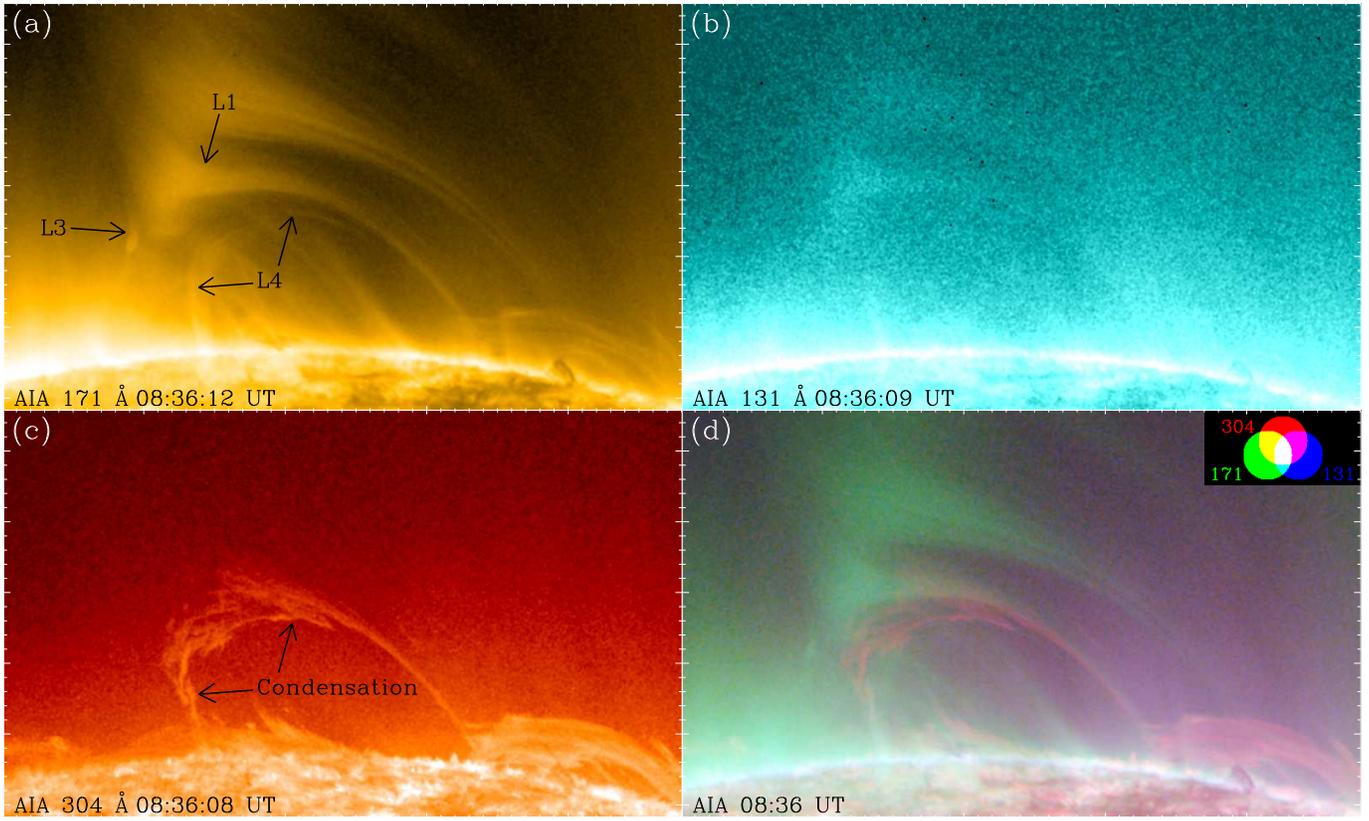}
\caption{Spatial relation between condensation downflows and loops. AIA 171\,\AA~(a), 131\,\AA~(b), and 304\,\AA~(c) images, and their composite (d), with the same FOV as in Figures\,\ref{f:reconnection}((a)-(c)) and \ref{f:downflow}((a)-(b)). (An animation of this figure is available.)
\label{f:animation}}
\end{figure}

Since $\sim$06:40 UT, the condensation flows along the right leg of loops L4, see Figure\,\ref{f:downflow}(b). A time-slice of AIA 304\,\AA~images is made along the blue line GH in Figure\,\ref{f:downflow}(b), and displayed in Figure\,\ref{f:downflow}(d). It indicates that the condensation first flows slowly, and then moves quickly, with a speed of 100 km\,s$^{-1}$ and an acceleration of 40 m\,s$^{-2}$. This flow pattern repeats itself frequently over the course of two hours. The spatial relation between condensation downflows and loops is displayed in Figure\,\ref{f:animation}.

\section{Summary and discussion}\label{sec:sum}

Employing SDO/AIA multi-wavelength images, we study the evolution of the MR between two sets of coronal loops L1 and L2, and the following condensation of hotter coronal plasma in the dip region of loops L1. The loops L1 show curved open structures with a dip in AIA 171\,\AA~images. They move downward and reconnect with the lower-lying closed loops L2. Newly-reconnected loops L3 and L4 form, and retract away from the MR region. Along with the MR, bright emission appears sequentially in the AIA 131\,\AA~and 304\,\AA~channels. The cooling and condensation of hotter coronal plasma takes place repeatedly in the dip of loops L1. The time delays between the peaks of the light curves of the AIA 171\,\AA~and 131\,\AA~channels, and the AIA 131\,\AA~and 304\,\AA~channels, are 30 and 50\,min, respectively. Due to the successive MR, the condensation, without support by lower-lying loops L1, flows toward the chromosphere first along the left leg, and then along the right leg, of the newly-reconnected loops L4.

\begin{figure}[ht!]
\centering
\includegraphics[width=0.5\textwidth]{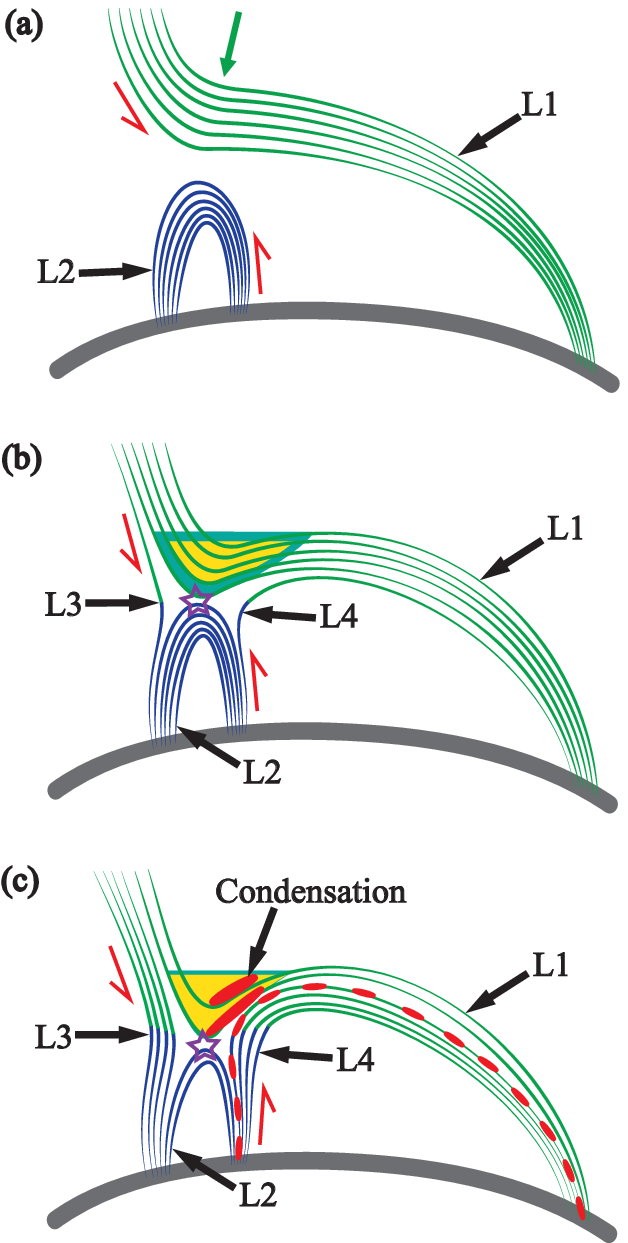}
\caption{Schematic diagrams of the MR and coronal condensation. In ((a)-(c)), the grey thick lines denote the solar limb. The green, blue, and green-blue lines separately show the magnetic field lines of loops L1, L2, L3, and L4, whose directions are marked by the red arrows. In (a), the green arrow denotes the dip of loops L1 and also the moving direction. In (b) and (c), the purple stars show the MR between loops L1 and L2. The green, yellow, and red patches indicate the AIA 171\,\AA, 131\,\AA, and 304\,\AA~plasma, respectively.
\label{f:cartoon}}
\end{figure}

According to the AIA extreme-ultraviolet (EUV) observations and the PFSS coronal magnetic field, schematic diagrams are provided in Figure\,\ref{f:cartoon} to describe the evolution of MR and coronal condensation. A set of open field lines (L1) come close to a region with closed field lines (L2), see Figure\,\ref{f:cartoon}(a), forming a dip in the former. Between them MR takes place, and newly-reconnected field lines (L3 and L4) form, see Figure\,\ref{f:cartoon}(b). Coronal material is gathered in the dip and cools, see Figure\,\ref{f:cartoon}(b). Through a loss of thermal equilibrium triggered by the MR, a condensation forms, and then slides down along the newly-formed field lines (L4), see Figure\,\ref{f:cartoon}(c).

The convergence of loops L1 and L2 toward an X-structure, and the retraction of newly-formed loops L3 and L4 away from the X-structure show clear observational evidence of MR occurring in the X-structure. The MR lasts for $\sim$22\,hr, much longer than any other MR events abundantly encountered on the Sun, which typically last only a few minutes, such as explosive events, EUV bursts, and jets, to a few hours, such as in flares. Both loops L1 and L2 are rooted in quiet Sun regions, with weaker magnetic field. Because much less magnetic flux reconnects in a much longer time, one can imagine that the MR rate here is much smaller, consistent with the observations that no associated flaring activity is detected during the MR in our case. 

Large coronal condensations are rarely observed. In this study, the coronal plasma cools from the AIA 171\,\AA~channel, rather than other AIA higher temperature channels, e.g. 193\,\AA~ \citep{liu12} and 211\,\AA~\citep{berg12}. The time delay between the AIA 171\,\AA~and 304\,\AA~channels is 80\,min, similar to \citet{vash15}, but less than \citet{liu12} and \citet{berg12}. Moreover, the AIA 131\,\AA~channel is employed to better show the cooling process of coronal plasma.

Condensations of plasma along open field lines are found. So far coronal rain has been reported to occur only along preexisting structures that are magnetically closed \citep{anto01,schr01}. When considering the AIA 304\,\AA~images alone, the condensations here would also resemble that occurring in closed loops. However, the source region for the condensation is the magnetically open structure L1 that forms a dip where the condensation takes place. In the later stages the condensation rains down not only along the leg of that open structure L1, but also through the MR region and down along the leg of the newly-reconnected field lines L4. This provides a new and alternative mechanism for the formation of coronal rain away from the magnetically closed regions that is initiated by MR. 

In traditional models, the thermal evolution, i.e. condensation, and the evolution of the magnetic field are treated separately \citep{mull03,xia16}. Mostly, it is assumed that the condensations form through a loss of equilibrium between heating and cooling. If the field line is stretched in the horizontal direction, and the condensation contains sufficient mass, then the field line would form a dip and plasma will remain in the lower density corona within the dip \citep{karp01}. Alternatively, MR can cause a helical structure with numerous dips during the prominence formation \citep{pneu83,kane15}. Considering the energy balance along these structures, one then gets condensations and plasma is trapped, again, in these dips \citep{pneu83,kane15,kane17}. In this Letter, both dips and condensations in the dips are formed, and they would resemble the prominence formation models \citep{pneu83,kane15}. However, the condensation remains for only a short time, and then rains down to the chromosphere when the dips are broken by the successive MR. In our study, we show that the MR and the condensation cannot be treated separately, but that plasma condensation naturally arises during the MR process.

\acknowledgments

The authors are indebted to the SDO team for providing the data. The work is supported by the National Foundations of China (11673034, 11533008, 11790304, and 11773039), and Key Programs of the Chinese Academy of Sciences (QYZDJ-SSW-SLH050). L. C. received funding from the European Union's Horizon 2020 research and innovation programme under the Marie Sklodowska-Curie grant agreement No. 707837. Project supported by the Specialized Research Fund for Shandong Provincial Key Laboratory.



\begin{thebibliography}{}

\bibitem[Antolin et al. (2001)]{anto01} Antolin, P., Vissers, G., \& Rouppe van der Voort, L.\ 2001, \solphys, 280, 457
\bibitem[Berger et al. (2012)]{berg12} Berger, T., Liu, W., \& Low, B.\ 2012, \apj, 758, L37
\bibitem[Huang et al. (2018)]{huang18} Huang, Z., Mou, C., Fu, H., et al.\ 2018, \apj, 853, L26
\bibitem[Innes et al. (2003)]{inne03} Innes, D., McKenzie, D., Wang, T.\ 2003, \solphys, 217, 267
\bibitem[Kaneko \& Yokoyama (2015)]{kane15} Kaneko, T., \& Yokoyama, T.\ 2015, \apj, 806, 115
\bibitem[Kaneko \& Yokoyama (2017)]{kane17} Kaneko, T., \& Yokoyama, T.\ 2017, \apj, 845, 12
\bibitem[Karpen et al. (2001)]{karp01} Karpen, J., Antiochos, S., Hohensee, M., Klimchuk, J., \& MacNeice, P.\ 2001, \apj, 553, L85
\bibitem[Kumar \& Cho (2013)]{kuma13} Kumar, P., \& Cho, K.\ 2013, \aap, 557, A115
\bibitem[Kwon et al. (2016)]{kwon16} Kwon, R., Vourlidas, A., \& Webb, D.\ 2016, \apj, 826, 94
\bibitem[Lemen et al. (2012)]{lemen12} Lemen, J., Title, A., Akin, D., et al.\ 2012, \solphys, 275, 17
\bibitem[Li et al. (2016c)]{li16c} Li, D., Ning, Z., \& Su, Y.\ 2016c, \apss, 361, 301
\bibitem[Li \& Zhang (2009)]{li09} Li, L., \& Zhang, J.\ 2009, \apj, 703, 877
\bibitem[Li et al. (2016a)]{li16a} Li, L., Zhang, J., Peter, H., et al.\ 2016a, NatPh, 12, 847
\bibitem[Li et al. (2016b)]{li16b} Li, L., Zhang, J., Su, J., \& Liu, Y.\ 2016b, \apj, 829, L33
\bibitem[Li et al. (2018)]{li18} Li, Y., Xue, J., Ding, M., et al.\ 2018, \apj, 853, L15
\bibitem[Lin \& Forbes (2000)]{lin00} Lin, J., \& Forbes, T.\ 2000, \jgr, 105, 2375
\bibitem[Liu et al. (2010)]{liu10} Liu, R., Lee, J., Wang, T., et al.\ 2010, \apj, 723, L28
\bibitem[Liu et al. (2012)]{liu12} Liu, W., Berger, T., \& Low, B.\ 2012, \apj, 745, L21
\bibitem[Malherbe et al. (1983)]{malh83} Malherbe, J., Priest, E., Forbes, T., \& Heyvaerts, J.\ 1983, \aap, 127, 153
\bibitem[Masuda et al. (1994)]{masu94} Masuda, S., Kosugi, T., Hara, H., Tsuneta, S., \& Ogawara, Y.\ 1994, \nat, 371, 495
\bibitem[McKenzie (2000)]{mcke00} McKenzie, D. E.\ 2000, \solphys, 195, 381
\bibitem[Mei et al. (2017)]{mei17} Mei, Z., Keppens, R., Roussev, I., \& Lin, J.\ 2017, \aap, 604, L7
\bibitem[M\"{u}ller et al. (2003)]{mull03} M\"{u}ller, D., Hansteen, V., \& Peter, H.\ 2003, \aap, 411, 605
\bibitem[Ning (2016)]{ning16} Ning, Z. J.\ 2016, \apss, 361, 22
\bibitem[Pesnell et al. (2012)]{pesn12} Pesnell, W., Thompson, B., \& Chamberlin, P.\ 2012, \solphys, 275, 3
\bibitem[Pneuman (1983)]{pneu83} Pneuman, G.\ 1983, \solphys, 88, 219
\bibitem[Priest \& Forbes (1986)]{prie86} Priest, E., \& Forbes, T.\ 1986,  \jgr, 91, 5579 
\bibitem[Priest \& Forbes (2000)]{prie00} Priest, E., \& Forbes, T.\ 2000,  Magnetic reconnection (Cambridge: Cambridge Univ. Press) 
\bibitem[Savage et al. (2012)]{sava12} Savage, S., McKenzie, D., Reeves, K.\ 2012, \apj, 747, L40
\bibitem[Schatten et al. (1969)]{scha69} Schatten, K., Wilcox, J., \& Ness, N.\ 1969, \solphys, 6, 442
\bibitem[Schrijver (2001)]{schr01} Schrijver, C.\ 2001, \solphys, 198, 325
\bibitem[Shibata (1999)]{shib99} Shibata, K.\ 1999, \apss, 264, 129
\bibitem[Smith \& Priest (1977)]{smith77} Smith, E., \& Priest, E.\ 1977, \solphys, 53, 25
\bibitem[Su et al. (2013)]{su13} Su, Y., Veronig, A., Holman, G., et al.\ 2013, NatPh, 9, 489
\bibitem[Takasao et al. (2012)]{taka12} Takasao, S., Asai, A., Isobe, H., Shibata, K.\ 2012, \apj, 745, L6
\bibitem[Tian et al. (2014)]{tian14} Tian, H., Li, G., Reeves, K., et al.\ 2014, \apj, 797, L14
\bibitem[Tsuneta et al. (1992)]{tsun92}  Tsuneta, S., Takahashi, T., Acton, L., et al.\ 1992, \pasj, 44, L211
\bibitem[Vashalomidze et al. (2015)]{vash15} Vashalomidze, Z., Kukhianidze, V., Zaqarashvili, T., et al.\ 2015, \aap, 577, A136
\bibitem[Xia \& Keppens (2016)]{xia16} Xia, C., \& Keppens, R.\ 2016, \apj, 823, 22
\bibitem[Xue et al. (2018)]{xue18} Xue, Z., Yan, X., Yang, L., et al.\ 2018, \apj, L4
\bibitem[Yan et al. (2018)]{yan18} Yan, X., Yang, L., Xue, Z., et al.\ 2018, \apj, 853, L18
\bibitem[Yang et al. (2018)]{yang18} Yang, B., Yang, J., Bi, Y., et al.\ 2018, \apj, 861, 135
\bibitem[Yang et al. (2017)]{yang17} Yang, L., Yan, X., Li, T., Xue, Z., \& Xiang, Y.\ 2017, \apj, 838, 131
\bibitem[Yang et al. (2015)]{yang15} Yang, S., Zhang, J., \& Xiang, Y.\ 2015, \apj, 798, L11
\bibitem[Yokoyama et al. (2001)]{yoko01} Yokoyama, T., Akita, K., Morimoto, T., Inoue, K., \& Newmark, J.\ 2001, \apj, 546, L69
\bibitem[Zheng et al. (2017)]{zheng17} Zheng, R., Chen, Y., Wang, B., Li, G., \& Xiang, Y.\ 2017, \apj, 840, 3

\end{thebibliography}
\end{document}